# Hamiltonian Formulation and Perturbations for Dust Motion around Cometary Nuclei

Yu Jiang[1,2]   Juergen Schmidt[3]   Hexi Baoyin[2]   Hengnian Li[2]   Junfeng Li[1]

1. School of Aerospace Engineering, Tsinghua University, Beijing 100084, China

2. State Key Laboratory of Astronautic Dynamics, Xi'an Satellite Control Center, Xi'an 710043, China

3. Astronomy and Astrophysics, University of Oulu, Finland

Y. Jiang (✉) e-mail: jiangyu_xian_china@163.com (corresponding author)

**Abstract**. In this paper we analyze the dynamical behavior of large dust grains in the vicinity of a cometary nucleus. To this end we consider the gravitational field of the irregularly shaped body, as well as its electric and magnetic fields. Without considering the effect of gas friction and solar radiation, we find that there exist grains which are static relative to the cometary nucleus; the positions of these grains are the stable equibria. There also exist grains in the stable periodic orbits close to the cometary nucleus. The grains in the stable equibria or the stable periodic orbits won't escape or impact on the surface of the cometary nucleus. The results are applicable for large charge dusts with small area-mass ratio which are near the cometary nucleus and far from the Solar. It is found that the resonant periodic orbit can be stable, and there exist stable non-resonant periodic orbits, stable resonant periodic orbits and unstable resonant periodic orbits in the potential field of cometary nuclei. The comet gravity force, solar gravity force, electric froce, magnetic force, solar radiation pressure, as well as the gas drag force are all considered to analyze the order of magnitude of these forces acting on the grains with different parameters. Let the distance of the dust grain relative to the mass centre of the cometary nucleus, the charge and the mass of the dust grain vary, respectively, fix other parameters, we clacluated the strengths of different forces. The motion of the dust grain depends on the area-mass ratio, the charge, and the distance relative to the comet's mass center. For a large dust grain (>1mm) close to the cometary nucleus which has a small value of area-mass ratio, the comet gravity is the largest force acting on the dust grain. For a small dust grain (<1mm) close to the cometary nucleus with large value of area-mass ratio, both the solar radiation pressure and the comet gravity are two major forces. If the a small dust grain which is close to the cometary nucleus have the large value of charge, the magnetic force, the solar radiation pressure, and the electric force are all major forces. When the large dust grain is far away from the cometary nucleus, the solar gravity and solar radiation pressure are both major forces.

**Key words**: Dust grains; Relative equilibria; Characteristic multipliers; Periodic orbits;

# 1. Introduction

Recently, several space missions were devoted to the exploration of comets, including Deep Impact and Rosetta (A'Hearn et al. 2005; Glassmeier et al. 2007). The NASA probe Deep Impact constrained the composition of comet 9P/Tempel by droping a projectile that collided with the nucleus of the comet at 5:52 UTC on July 4, 2005 (A'Hearn et al. 2005; Lisse et al. 2006). The Rosetta mission encountered comet 67P/Churyumov–Gerasimenko in 2014, finding that the nucleus consists of two lobes connected by a short neck (Capaccioni et al. 2015; Sierks et al. 2015). Either the two lobes represent a contact binary body, having formed about 4.5 billion years ago, or a single body has formed a gap by mass loss (Sierks et al. 2015).

Space missions to comets revive the interest in the dynamics of dust in the potential of a cometary nucleus. The spacecraft Stardust brought interstellar particles and grains from comet 81P/Wild2 back to Earth on 2006 (Brownlee et al. 2006). There were 256 grains found on the collector surface with sizes larger than 100μm and about 1200 grains larger than 1 μm (Burchell et al. 2008). The dust grains can be produced on active areas when cometary ices evaporate if the comet is close enough to the sun (Belton 2010). Oberc (1997) investigates only the ejection of secondary particles from larger aggregates already flying in the coma, not the direct ejection from the nucleus. For the latter, both electrostatic and centrifugal forces are likely of minor relevance. Fulle et al. (1997) discussed the sunward structure in the comet 19P/Borrelly's dust tail. The nucleus can have one or more active areas producing gas and dust (Combi et al. 2012; Yu et al. 2016). Belton (2010) discussed the active areas of 1P/Halley, 19P/Borrelly, 81P/Wild 2, and 9P/Tempel 1 respectively, and found there are 3 types of active areas. For Type I, $H_2O$ is sublimated through the porous mantle. For Type II, super-volatiles from the interior is persistent effused; while for Type III, super-volatiles from the interior is episodic released.

Richter and Keller (1995) studied the motion stability of dust particle with considering the weak solar radiation pressure force and the gravitational force of a spherical body. Liu et al. (2011) investigated the equilibrium points and periodic motion around a rotating cube. However, cometary nuclei possess irregular shapes (Stern 2003; Jiang et al. 2014; Wang et al. 2014; Jiang et al. 2015a). To understand the dynamical behavior of grains in the gravitational potential of an irregularly shaped body, several models for irregular nuclei have been investigated in the literature, including a straight segment (Romero et al 2004; Lindner et al. 2010; Najid et al. 2011), a triangular plate and a square plate (Blesa 2006), a homogeneous annular disk (Alberti and Vidal 2007; Fukushima 2010) and a circular ring (Najid et al. 2012). Hughes (2000) studied the final velocity of dust particles emitted by a nucleus, while Farnocchia et al. (2014) discussed the ejection velocity from comet C/2013 A1. Moreno et al. (2012) derived dust loss rates, ejection velocities, and power-law size distributions as functions of the heliocentric distance. Rotundi et al. (2015) reported the velocity distribution relative to comet 67P/Churyumov-Gerasimenko's body-fixed frame of grains as observed by the spacecraft Rosetta.

The equations of motion around the cometary nucleus were derived applying the Hill approximation (Scheeres and Marzari 2002). Using an expansion in Legendre polynomials for the potential of the nucleus, the gravitational potential diverges within the Brillouin sphere, and the spacecraft dynamics cannot be modeled close the surface (Takahashi et al. 2013; Wang et al. 2014). The characteristics of the interior gravity field are derived near the surface of the asteroid and comet for the purpose of small body proximity operations (Takahashi et al. 2013).

The dynamical behaviours around comets include equilibria and periodic orbits. The equilibrium points are the critical points of the effective potential. The physical characteristics of equilibrium points means that there exist a physical point in actual 3D space where one can put a particle such that the resultant external force of the massless particle relative to the body of the comet nucleus is zero. If an equilibrium points is stable, the particle will not move (relative to the nucleus). In the body-fixed frame, the equilibrium point is not the stable periodic orbit. However, relative to the inertia frame, the equilibrium point is a periodic orbit. The Floquet multiplyers of a periodic orbit mean the eigenvalues of monodromy matrix of the periodic orbit. The distribution of the Floquet multiplyers confirms the stability of the periodic orbit. The topological cases of a periodic orbit confirm its stable level. The grain may move in the stable periodic orbit relative to the comet, the Floquet multiplyers can help one to confirm the stability of the periodic orbit and point out if a periodic orbit can be possible for the grain. Wang et al. (2014) calculated the locations and stability of equilibrium points of comet 1P/Halley, 9P/Tempel 1, and 103P/Hartley 2, and found each of these three comets has five equilibrium points. Jiang et al. (2015b) found two different periodic orbits with different topological cases around comet 1P/Halley.

Jiang and Baoyin (2016) investigated the continuation of periodic orbit family around 1P/Halley, and pointed out that the collisions of Floquet multipliers may be maintained during the continuation.

This paper investigates the dynamical behavior of large dust grains in the vicinity of a cometary nucleus, including the types of local motion near equilibrium points, i.e. stable global motion, unstable global motion, and resonant motion. In addition to the irregular gravitational field, electric and magnetic fields are considered. The effect of gas friction and solar radiation are neglected in a Hamiltonian approach to describe the dust's motion. The approximations are applicable for dusts with small area-mass ratio ($<1.0\times10^5 m^2 kg^{-1}$), large charge ($>1.0\times10^{-18}$C), short distances of the dust grain to the mass center of the cometary nucleus (<10km), as well as long distances of the dust grain to the Solar (>3-4 AU). The Lorentz force is thought to become relevant only under very specific conditions far from the Sun (Kramer et al., 2014).

Linearised equations of motion for the dust particles around equilibrium points in the potential of the cometary nucleus are derived. The characteristic equation of the dust grains' motion around equilibrium points is also presented. Furthermore, a corollary for a sufficient condition of the linear stability is presented and proved. An identical equation with regard to the number of non-degenerate equilibria in the gravitational field, electric field, and magnetic field of the cometary nucleus is presented and proved. We find that the number of non-degenerate equilibria for the dust grain only varies in pairs, and the number of non-degenerate equilibrium points is an odd number.

We identify five equilibrium points in the combined gravitational, electric, and magnetic fields of the nucleus of comet 1P/Halley. Positions and eigenvalues of these equilibrium points are calculated. Although Rosetta found indication of grains in orbit about the nucleus in the early phase of the mission (Rotundi et al., 2015), these did not pose a threat to the mission. The biggest problem was caused by a huge number of large grains on un-bound orbits that confused the software of the startracker sensors, disturbing the on-board navigation. There exist stable non-resonant periodic orbits, stable resonant periodic orbits and unstable resonant periodic orbits in the combined gravitational and electric field of comet 1P/Halley. The periodic orbits with the 1:1, 1:2, and 1:8 resonances are discussed.

Perturbations acting on the dust grain are also considered. To compare different forces, some parameters are fixed and let other parameters change. The comet gravity is the biggest force when the dust grain is near the cometary nucleus. The comet gravity and the electric and magnetic forces decrease rapidly while the distance from the dust grain to the mass center of the cometary nucleus is increasing. When the dust grain is far away from the cometary nucleus, the solar gravity and the solar radiation pressure are the major forces.

**2. Dynamical Equation and Effective Potential**

Consider a dust grain which orbits around the cometary nucleus. Denoting with **r** the radius vector from the nucleus's centre of mass to the dust grain, the first and second time derivatives of **r** are expressed with respect to the body-fixed coordinate system of the nucleus. The reference coordinate system used throughout this paper is the body-fixed frame. Denoting with $\boldsymbol{\omega}$ the angular velocity vector of the cometary nucleus relative to the inertial space, the generalised momentum of the dust grain is $\mathbf{p}=(\dot{\mathbf{r}}+\boldsymbol{\omega}\times\mathbf{r})$, and the generalised coordinate is $\mathbf{q}=\mathbf{r}$. Further, we denote with $U(\mathbf{r})$ the gravitational potential of the cometary nucleus, and with $\mathbf{E}(\mathbf{r})$ and $\mathbf{B}(\mathbf{r})$ the electric field intensity and the magnetic flux density of the cometary nucleus, respectively.

The gravitational fields and shape models of cometary nuclei can be calculated using the polyhedral model (Werner 1994; Werner and Scheeres 1997) using data from radar observations or spacecraft images (Stooke 2002). Spacecraft images provide excellent data for shape models. With the polyhedron method, the gravitational potential (Werner and Scheeres 1997) of a cometary nucleus can be represented by

$$U = \frac{1}{2} G\sigma \sum_{e \in edges} \mathbf{r}_e \cdot \mathbf{E}_e \cdot \mathbf{r}_e \cdot L_e - \frac{1}{2} G\sigma \sum_{f \in faces} \mathbf{r}_f \cdot \mathbf{F}_f \cdot \mathbf{r}_f \cdot \omega_f, \qquad (1)$$

where $G=6.67 \times 10^{-11}$ m$^3$kg$^{-1}$s$^{-2}$ is the gravitational constant and $\sigma$ is the cometary nucleus' density; $L_e$ is a factor of integration and $\omega_f$ is the signed solid angle; $\mathbf{r}_e$ and $\mathbf{r}_f$ are body-fixed vectors from the field point to fixed points on the edges and the faces, respectively; $\mathbf{E}_e$ and $\mathbf{F}_f$ are dyads representing geometric parameters of edges and faces, respectively. The electric field intensity of the cometary nucleus can be calculated from

$$\mathbf{E}(\mathbf{r}) = \frac{1}{4\pi\varepsilon_0} \int_V \frac{\rho(\mathbf{r}_1)}{r_s^3} \mathbf{r}_s dV(\mathbf{r}_1) = -\nabla\phi(\mathbf{r}), \qquad (2)$$

where $\mathbf{r}_s = \mathbf{r} - \mathbf{r}_1$, $\varepsilon_0 = 8.854 \times 10^{-12} \, \text{F} \cdot \text{m}^{-1}$ is the vacuum permittivity, and $\rho(\mathbf{r}_1)$ and $\phi$ are the charge density and electrostatic potential of the cometary nucleus, respectively. The magnetic vector potential is

$$\mathbf{A}(\mathbf{r}) = \frac{\mu_0}{4\pi} \int_V \frac{J(\mathbf{r}_1)}{r_s} dV(\mathbf{r}_1), \quad (3)$$

where $\mu_0$ is the vacuum permeability. Then the magnetic flux density of the cometary nucleus can be calculated by

$$\mathbf{B}(\mathbf{r}) = \nabla \times \mathbf{A}(\mathbf{r}). \quad (4)$$

Let $Q$ be the electrical charge of the dust grain. Then the total gravitational and electromagnetic force felt by the particle is given by

$$\mathbf{f} = -\nabla U + Q(\mathbf{E} + \mathbf{v} \times \mathbf{B}), \quad (5)$$

where $\mathbf{v} = \dot{\mathbf{r}}$. In a frame that rotates with the comet nucleus, the equation of motion (Jiang 2015; Jiang and Baoyin 2016) for the dust grain reads

$$\ddot{\mathbf{r}} + 2\boldsymbol{\omega} \times \dot{\mathbf{r}} + \boldsymbol{\omega} \times (\boldsymbol{\omega} \times \mathbf{r}) = -\nabla U + Q(\mathbf{E} + \mathbf{v} \times \mathbf{B}). \quad (6)$$

Then the integral of the relative energy is

$$H = -\frac{\mathbf{p} \cdot \mathbf{p}}{2} + U(\mathbf{q}) + \mathbf{p} \cdot \dot{\mathbf{q}} + Q\phi(\mathbf{q}) = \frac{1}{2}(\mathbf{p} - \boldsymbol{\omega} \times \mathbf{q}) \cdot (\mathbf{p} - \boldsymbol{\omega} \times \mathbf{q}) + V(\mathbf{q}) + Q\phi(\mathbf{q}). \quad (7)$$

The nucleus of 1P/Halley is an irregular, potato-shaped body (Sagdeev et al. 1986). The estimated bulk density of the nucleus is $0.6 \, \text{g} \cdot \text{cm}^{-3}$ (Sagdeev et al. 1988), its rotational period is 52.8 h, and the overall dimension is $16.831 \times 8.7674 \times 7.7692$ km (Peale and Lissauer 1989; Stooke 2002). Figure 1 shows the 3D shape represented with the polyhedral model (Werner 1994; Werner and Scheeres 1997) using shape data from Stooke (2002).

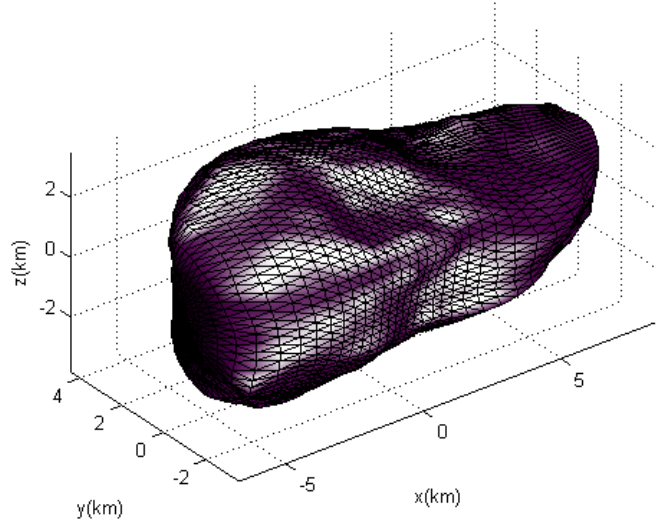

Fig. 1. The 3D shape of 1P/Halley. The shape was built with 5040 faces using the polyhedron model. Convex surfaces are plotted in light color.

Figure 2 shows a contour plot of the effective gravitational potential (see Appendix A) of 1P/Halley calculated from this model. 1P/Halley has five equilibrium points (see Appendix A), four of them lying outside the body and one inside the body. The positions of these five equilibrium points in the body-fixed frame are listed in Table 1; among them, E5 is inside the comet. Table 2 lists the eigenvalues of the equilibrium points. The form of the eigenvalues for the equilibrium points E1 and E3 is $\lambda_{1,2} = \pm i\beta_1 (\beta_1 > 0), \lambda_{3,4} = \pm i\beta_2 (\beta_2 > 0)$ and $\lambda_{5,6} = \pm \alpha_1 (\alpha_1 > 0)$, while for the equilibrium points E2 and E4 the eigenvalues have the form $\lambda_{1,2} = \pm i\beta_1 (\beta_1 > 0)$, $\lambda_{3,4} = \pm i\beta_2 (\beta_2 > 0), \lambda_{5,6} = \pm i\beta_3 (\beta_3 > 0)$. The motion of a dust grain relative to the equilibrium points E1 or E3 of 1P/Halley can be expressed as

$$\begin{cases} \xi = \mathbf{C}_\xi \left[ e^{\alpha_1 t}, e^{-\alpha_1 t}, \cos\beta_1 t, \sin\beta_1 t, \cos\beta_2 t, \sin\beta_2 t \right]^\mathrm{T} \\ \eta = \mathbf{C}_\eta \left[ e^{\alpha_1 t}, e^{-\alpha_1 t}, \cos\beta_1 t, \sin\beta_1 t, \cos\beta_2 t, \sin\beta_2 t \right]^\mathrm{T} \\ \zeta = \mathbf{C}_\zeta \left[ e^{\alpha_1 t}, e^{-\alpha_1 t}, \cos\beta_1 t, \sin\beta_1 t, \cos\beta_2 t, \sin\beta_2 t \right]^\mathrm{T} \end{cases} \quad (8)$$

while the motion of the dust grain relative to the equilibrium points E2 or E4 follows a quasi-periodic orbit, which is expressed as

$$\begin{cases} \xi = \mathbf{C}_\xi \left[\cos\beta_1 t, \sin\beta_1 t, \cos\beta_2 t, \sin\beta_2 t, \cos\beta_3 t, \sin\beta_3 t\right]^T \\ \eta = \mathbf{C}_\eta \left[\cos\beta_1 t, \sin\beta_1 t, \cos\beta_2 t, \sin\beta_2 t, \cos\beta_3 t, \sin\beta_3 t\right]^T \\ \zeta = \mathbf{C}_\zeta \left[\cos\beta_1 t, \sin\beta_1 t, \cos\beta_2 t, \sin\beta_2 t, \cos\beta_3 t, \sin\beta_3 t\right]^T \end{cases}. \quad (9)$$

Here, $\mathbf{C}_\xi$, $\mathbf{C}_\eta$, $\mathbf{C}_\zeta$ are $1\times 6$ vectors. The motion is unstable in the vicinity of equilibrium points E1 and E3, and linearly stable in the vicinity of equilibrium points E2 and E4.

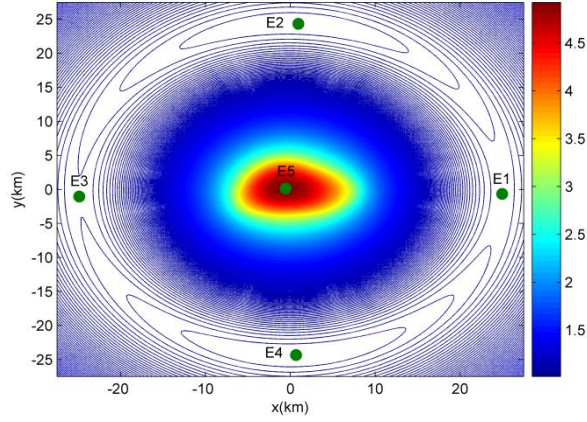

Fig. 2. Contour plot of the effective gravitational potential for the nucleus of 1P/Halley. The unit of the effective potential is $m^2 s^{-2}$

Table 1 Positions of the equilibrium points for dust motion in the effective gravitational potential of comet 1P/Halley

| Equilibrium Points | x (km) | y (km) | z (km) |
|---|---|---|---|
| E1 | 24. 946 | -0.662 | 0.004 |
| E2 | 0.945 | 24.341 | -0.001 |
| E3 | -24.856 | -1.052 | 0.005 |
| E4 | 0.675 | -24.322 | 0.000 |
| E5 | -0.577 | 0.142 | 0.008 |

Table 2 Eigenvalues of the equilibrium points for dust motion in the effective gravitational potential of comet 1P/Halley

| $\times 10^{-4} s^{-1}$ | $\lambda_1$ | $\lambda_2$ | $\lambda_3$ | $\lambda_4$ | $\lambda_5$ | $\lambda_6$ |
|---|---|---|---|---|---|---|
| E1 | 0.340i | -0.340i | 0.346i | -0.346i | 0.131 | -0.131 |

| E2 | 0.124i | -0.124i | 0.306i | -0.306i | 0.331i | -0.331i |
| E3 | 0.338i | -0.338i | 0.343i | -0.343i | 0.114 | -0.114 |
| E4 | 0.139i | -0.139i | 0.299i | -0.299i | 0.331i | -0.331i |
| E5 | 4.739i | -4.739i | 4.448i | -4.448i | 2.880i | -2.880i |

Assume the dust grain has $N$ relative equilibria, which include equilibria inside the body of the nucleus, and denote $\sigma_j(E_k)$ as the $j$ th eigenvalue of the $k$ th equilibrium point. Then we have the following theorem constraining the number of non-degenerate equilibria in the combined gravitational, electric, and magnetic fields of a cometary nucleus.

**Theorem 1.** Eigenvalues of all the relative equilibria for the dust grain satisfy the identical equation $\sum_{k=1}^{N}\left[\operatorname{sgn}\prod_{j=1}^{6}\sigma_j(E_k)\right]=\sum_{j=1}^{N}\left[\operatorname{sgn}\left(\det\left(\nabla^2 V\right)\right)\right]=const$.

The proof of theorem 1 can be seen in Appendix B.

If an equilibrium point has at least one eigenvalue equal to zero, then the equilibrium point is called a degenerate equilibrium point. Conversely, if all the eigenvalues of an equilibrium point are non-zero, the equilibrium point is called a non-degenerate equilibrium point. For the non-degenerate equilibrium points of the dust motion we have the following

**Corollary 2.** The number of non-degenerate equilibria in the gravitational field, electric field, and magnetic field of the cometary nucleus only varies in pairs.

In addition, for the degenerate equilibrium points of dust motion, we know that the change of the equilibrium type belongs to one of the following paths: (1) annihilate; (2) transforms to arbitrary number of degenerate equilibrium points; (3) transforms to even number of non-degenerate equilibrium points; (4) transforms to arbitrary number of degenerate equilibrium points and even number of non-degenerate equilibrium points. For instance, the degenerate equilibrium point perhaps varies to 4 non-degenerate equilibrium points and 5 degenerate equilibrium points, or varies to 2 non-degenerate equilibrium points, or vanishes. It is impossible for the degenerate equilibrium point of the dust grains in these three fields of cometary nucleus varies to odd number of non-degenerate equilibrium points.

**Corollary 3.** The number of non-degenerate equilibrium points in the gravitational field, electric field, and magnetic field of the cometary nucleus is an odd number. It can be 1, 3, 5, 7, 9, …, etc.

## 3. Stability, Bifurcation, and Resonance of Periodic Orbits

In this section, the stability properties, bifurcations, and resonances of periodic orbits of dust grains in the irregular gravitational field, electric field, and magnetic field of a cometary nucleus are analyzed. First, we present the topological classification for stable periodic and unstable periodic orbits; then conditions for occurrence of bifurcations are discussed; furthermore, resonances are investigated, which take place when the orbital period of the grain and the rotation period for the nucleus form an integer ratio. In addition, several numerical results about periodic orbits are presented and analyzed.

### 3.1 Stability of Periodic Orbits

We consider the set $S_p(T)$ of periodic solutions around a cometary nucleus, which can be expressed as $\mathbf{z}(t) = \mathbf{f}(t, \mathbf{z}_0)$, $\mathbf{f}(0, \mathbf{z}_0) = \mathbf{z}_0$. Let $S_p(T)$ be the set of periodic orbits with the period $T$. Denote the matrix $\nabla \mathbf{f} := \dfrac{\partial \mathbf{f}(\mathbf{z})}{\partial \mathbf{z}}$, then the state transition matrix of the periodic orbit $p \in S_p(T)$ is

$$\Phi(t) = \int_0^t \nabla \mathbf{f}(p(\tau))d\tau, \tag{10}$$

and the monodromy matrix of the periodic orbit is

$$M = \Phi(T). \tag{11}$$

Eigenvalues of the matrix $M$ are the characteristic multipliers of the periodic orbit $p$. The monodromy matrix $M$ is a symplectic matrix, so if $\lambda$ is a characteristic multiplier of the periodic orbit, then $\lambda^{-1}$, $\bar{\lambda}$, and $\bar{\lambda}^{-1}$ are also characteristic multipliers of the periodic orbit. Considering the topological classification for the periodic orbit of six characteristic multipliers on the complex plane, there are five classes of stable periodic orbits and six classes of unstable periodic orbits. All the characteristic multipliers of a stable periodic orbit lie on the unit cycle, while an unstable periodic orbit has characteristic multipliers that don't lie on the unit cycle. Figure 3 shows the topological classification for the stable periodic orbit of six characteristic multipliers on the complex plane, while Figure 4 shows it for the unstable periodic orbit.

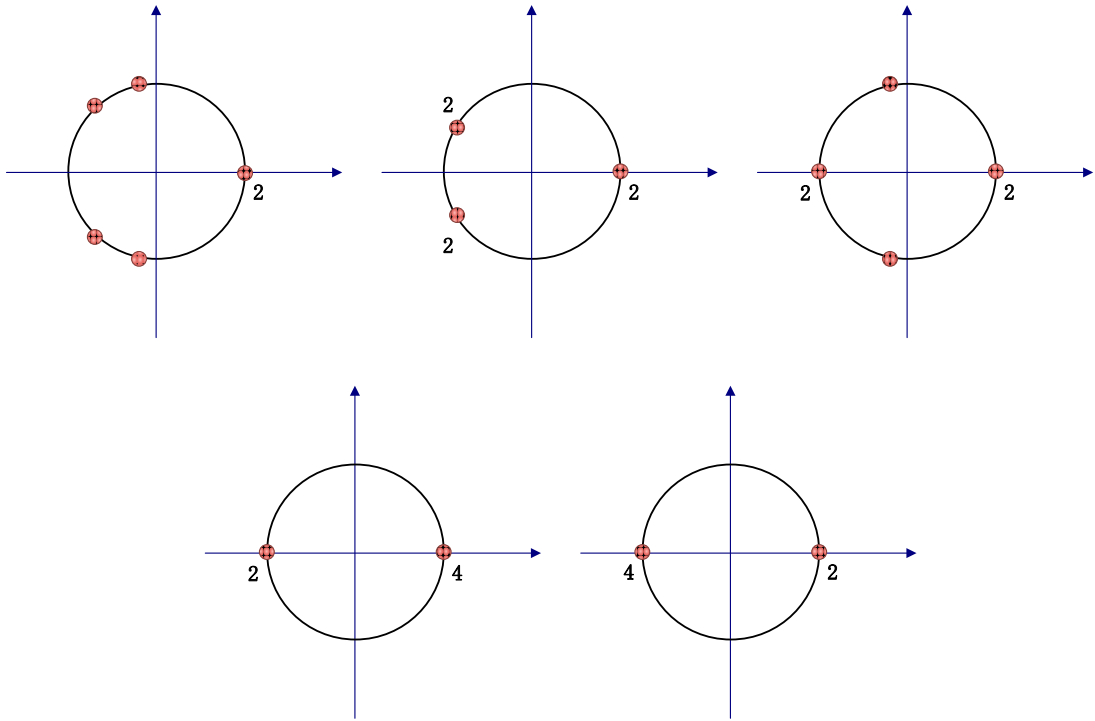

Fig. 3. The topological classification for the stable periodic orbits of six characteristic multipliers on the complex plane

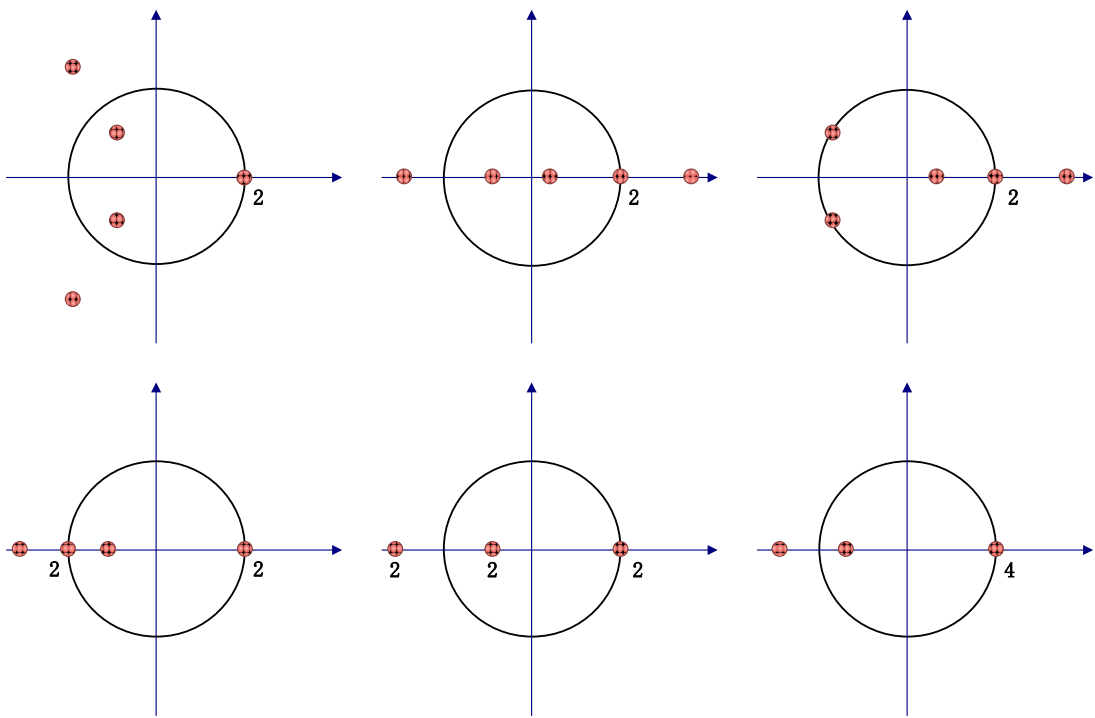

Fig. 4. The topological classification for the unstable periodic orbits of six characteristic multipliers on the complex plane

**3.2 Bifurcations of Orbit Families for Dust Grains**

For a long time, the orbits of the grains will vary. During the variety of the orbits, if the stability of the orbit remains unchanged, the grains will moves on the orbit; otherwise, the grains will escape or impact on the surface of the cometary nucleus. The bifurcations is related to the stability of the orbits, thus the analysis of the bifurcations of the orbits can help one to understand the variety of the motion state of the grains, including the escape and the impact. In this section, we first discuss bifurcations of orbit families in the gravitational field of cometary nuclei. Bifurcations of periodic orbits in the potential of a cometary nucleus occur when the topological cases and stability of the periodic orbits vary. There are four kinds of bifurcations, the tangent bifurcation, the period-doubling bifurcation, the Neimark-Sacker bifurcation and the real saddle bifurcation. If two characteristic multipliers coalesce at $\lambda=1$ we have a tangent bifurcation and if they coalesce at $\lambda=-1$ a period-doubling bifurcation; in addition, if two complex conjugate pairs of characteristic multipliers collide, we have the case of a Neimark-Sacker bifurcation or a real saddle bifurcation. A collision point on the unit circle ($\lambda\neq\pm1$) corresponds to the Neimark-Sacker bifurcation while a collision point on the real axis ($\lambda\neq\pm1$) corresponds to the real saddle bifurcation. First we discuss the bifurcations of arbitrary orbit families, then we discuss the bifurcations of periodic orbit families. Figure 5 shows appearance of the period-doubling bifurcation. In Figure 5, one can see that before the bifurcation, two characteristic multipliers on the unit circle approach each other, and coincide at -1 when the bifurcation occurred, after the bifurcation, these two characteristic multipliers enter the real axis.

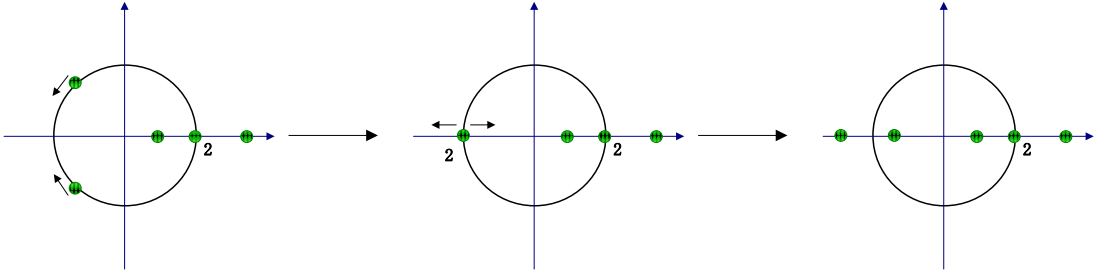

Fig. 5 Appearance of the period-doubling bifurcation, distributions of 6 characteristic multipliers on the complex plane are shown, the radius of the circle is 1.

Now we focus on bifurcations of arbitrary orbit families for charged dust grains in the cometary nucleus' irregular gravitational field, electric field, and magnetic field. The dust grains are supposed to have electrical charge $Q$. From Eq. (11), one can see that the characteristic polynomial of an arbitrary orbit relative to the fixed frame of the nucleus is

$$P(\lambda)=|\lambda E - M| = \lambda^6 + k_1\lambda^5 + k_2\lambda^4 + k_3\lambda^3 + k_2\lambda^2 + k_1\lambda + 1. \qquad (12)$$

$P(\lambda)$ satisfies $P(\lambda^{-1}) = \lambda^{-6}P(\lambda)$ because $M$ is a symplectic matrix. A tangent bifurcation occurs when $P(1)=0$, while a period-doubling bifurcation occurs when $P(-1)=0$. Thus, the condition for the tangent bifurcation is $2k_1 + 2k_2 + k_3 + 2 = 0$ while for the period-doubling bifurcation it reads $-2k_1 + 2k_2 - k_3 + 2 = 0$. $P(\lambda)$ has multiple roots when a Neimark-Sacker bifurcation or real saddle bifurcation occurs.

We denote $\rho = \lambda + \frac{1}{\lambda}$. Then Eq. (12) becomes

$$\rho^3 + k_1\rho^2 + (k_2 - 3)\rho + k_3 - 2k_1 = 0. \tag{13}$$

Denoting $\tilde{\rho} = \rho - \frac{k_1}{3}$, we find that Eq. (13) reduces to

$$\tilde{\rho}^3 + 3p\tilde{\rho} + 2q = 0, \tag{14}$$

where $3p = -\frac{k_1^2}{3} + k_2 - 3$ and $2q = 2\left(\frac{k_1}{3}\right)^3 - \frac{k_1(k_2-3)}{3} + k_3 - 2k_1$. The discriminant of Eq. (14) is

$$\Delta = q^2 + p^3 = \left[\left(\frac{k_1}{3}\right)^3 - \frac{k_1(k_2-3)}{6} + \frac{k_3 - 2k_1}{2}\right]^2 + \left[\frac{k_2-3}{3} - \frac{k_1^2}{9}\right]^3 = 0. \tag{15}$$

Next we consider the bifurcations of periodic orbit families for uncharged dust grains. The characteristic polynomial of the periodic orbit is

$$Q(\lambda) = |\lambda E - M| = (\lambda - 1)^2 (\lambda^4 + l_1\lambda^3 + l_2\lambda^2 + l_1\lambda + 1). \tag{16}$$

The condition for the tangent bifurcation is $2l_1 + l_2 + 2 = 0$ while for the period-doubling bifurcation it reads $2l_1 - l_2 - 2 = 0$. $Q(\lambda)$ has multiple roots when the Neimark-Sacker bifurcation or the real saddle bifurcation occurs, which satisfies $l_1^2 - 4l_2 + 8 = 0$.

### 3.3 Resonant Orbit Families for Dust Grains

Resonant periodic orbits arise if the orbital period for the dust grain and the rotation period for the cometary nucleus are in an integer ratio. Let $T_n$ be the rotation period of the nucleus, $T_o$ be the orbital period of the dust grain. If $\frac{T_o}{T_n} = \frac{m_o}{m_n}$, then $m_o : m_n$ is the resonant ratio. In the vicinity of a cometary nucleus, there exist not only unstable resonant periodic orbits, but also stable resonant periodic orbits.

The dust aggregates can be ejected from the surface of the cometary nucleus by electrostatic and rotational ejection (Oberc 1997; Crifo et al. 2005); the action of non-gravitational forces makes the ejection velocity size-dependent (Oberc 1997; Molina et al. 2008). Banaszkiewicz et al. (1990) modeled the gravitational potential by a triaxial ellipsoid and calculated the orbits of dust around cometary nuclei. Oberc (1997) considered both, electrostatic and rotational ejection and analysed the relationship between the ejection velocity and the aggregatse size. Molina et al. (2008) modeled the gravitational potential of comet 46P/Wirtanen by a sphere and considered the effect of gas drag. They computed the orbits of the largest grain ejected from the surface of the nucleus with a size of 5 cm. Comet 1P/Halley has both electric field and magnetic field (Horanyi and Mendis 1986; Delva et al. 2014).

The gravitational model we use here is more complex than models used in previous studies. To this end, we calculate the irregular gravitational field of 1P/Halley is by the polyhedral method (Werner 1994; Werner and Scheeres 1997), using data from radar observations (Neese 2004). The electrical field is approximated as the one generated by a point charge (Hartzell 2013). For this configuration the periodic orbits are calculated for 1P/Halley. Lacking knowledge of the precise value for the charge of 1P/Halley (Horanyi and Mendis 1986; Delva et al. 2014), we assume for the comet and the dust grain charges of $1.0 \times 10^{-2} C$ and $2.8 \times 10^{-13} C$, respectively. The magnetic field is neglected here. Figure 6 shows 5 periodic orbits for dust grains found around the nucleus of 1P/Halley.

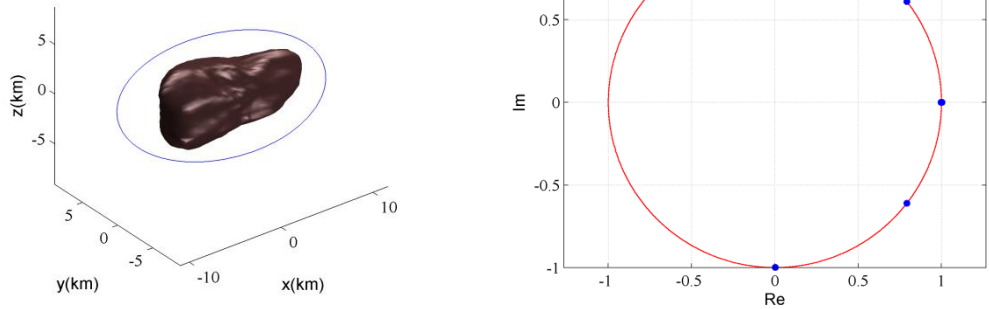

Fig. 6.a. A stable, non-resonant periodic orbit around the nucleus of 1P/Halley with period 9.51893 h. Right panel: The distribution of characteristic multipliers in the complex plane for this periodic orbit.

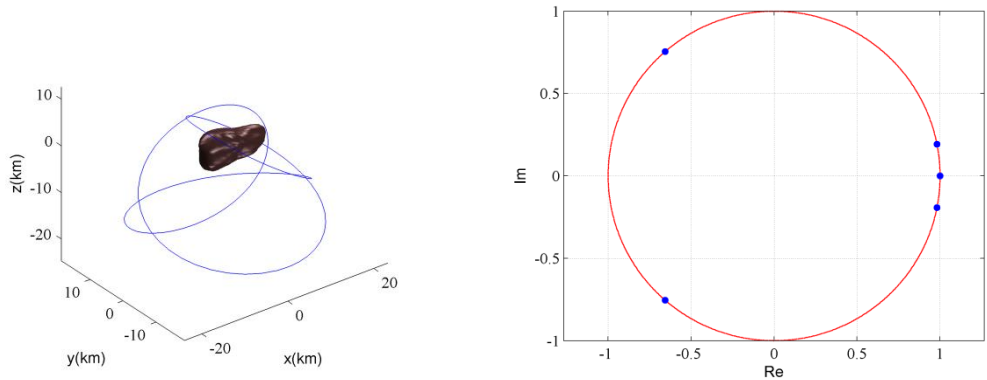

Fig. 6.b. A stable, non-resonant periodic orbit around the nucleus of 1P/Halley with period 58.8557 h. Right panel: The distribution of characteristic multipliers in the complex plane for the periodic orbit.

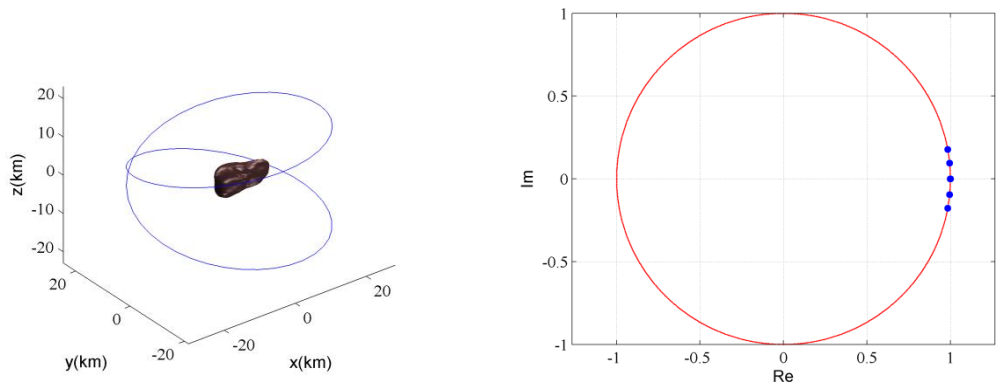

Fig. 6.c. A stable 1:1 resonant periodic orbit around the nucleus of 1P/Halley with period 53.4299 h and distribution of characteristic multipliers in the complex plane for the periodic orbit.

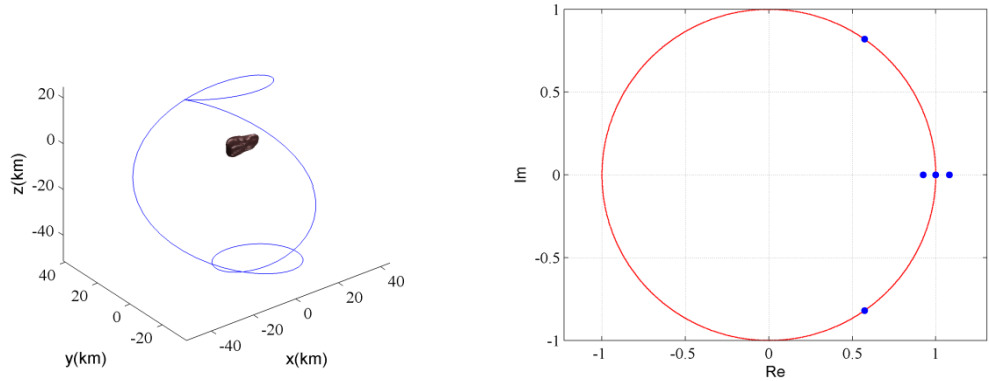

Fig. 6.d. An unstable resonant periodic orbit around the nucleus of 1P/Halley with period 105.7575 h and distribution of characteristic multipliers in the complex plane for the periodic orbit.

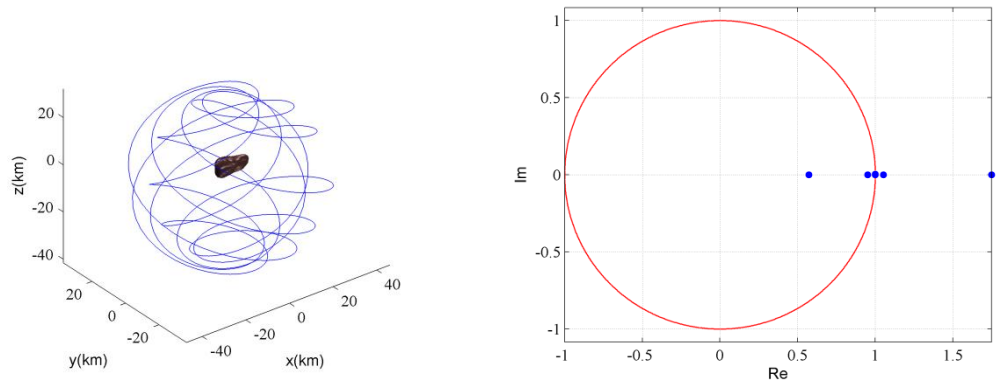

Fig. 6.e. An unstable resonant periodic orbit around the nucleus of 1P/Halley with period 423.022 h, and distribution of characteristic multipliers in the complex plane for the periodic orbit

Figure 6.a shows a periodic orbit around the nucleus of 1P/Halley with period 9.51893 h as well as the corresponding distribution of characteristic multipliers in the complex plane. The ratio of the orbital period for the dust grain and the rotation period for the cometary nucleus equals 0.1803. This periodic orbit is stable and non-resonant, which is one of the cases shown in Figure 3. All the characteristic multipliers are on the unit circle. Figure 6.b shows another stable, non-resonant periodic orbit with period 58.8557 h, the ratio of period is in this case 1.115. The configuration of characteristic multipliers in the complex plane shows that this is one of the cases shows in Figure 3.

Resonant periodic orbits in the gravitational field of a cometary nucleus can also be stable. Figure 6.c shows such a stable 1:1 resonant periodic orbit around 1P/Halley with period 53.4299 h. The topological classification for this orbit is one of the cases shown in Figure 3. The 1:1 resonant periodic orbit presented here is different from the periodic orbit near an equilibrium point; because there are three families of periodic orbits near the equilibrium point E2; the first family of periodic orbits near the equilibrium point E2 have period about 52.7h, corresponding to the eigenvalue $0.331i \times 10^{-4}$; the second one near the equilibrium point E2 have period about 57.0h, corresponding to the eigenvalue $0.306i \times 10^{-4}$; the third one have period about 140.8h corresponding to the eigenvalue $0.124i \times 10^{-4}$.

Figure 6.d shows an unstable 1:2 resonant periodic orbit around 1P/Halley with period 105.7575 h. The distribution of characteristic multipliers for this periodic orbit shows that this corresponds to one of the cases shown in Figure 4.

Figure 6.e shows an unstable 1:8 resonant periodic orbit around 1P/Halley with period 423.022 h. The distribution of characteristic multipliers for this periodic orbit shows that the topological classification for this unstable resonant periodic orbit corresponds to one of the cases shown in Figure 4.

The application to comet 1P/Halley implies that stable and non-resonant periodic orbits, stable 1:1 resonant periodic orbits, as well as unstable resonant periodic orbits can all simultaneously exist around the same cometary nucleus.

## 4. Solar Gravity Force and Solar Radiation Pressure

For a dust grain, there exist several kinds of forces on it, including the gravity force caused by the irregularly shaped cometary nucleus, the electric and magnetic force, the solar gravity force and the solar radiation pressure. The Finson-Probstein model only has two forces, the solar gravity force and the solar radiation pressure (Finson and Probstein 1968; Kramer et al. 2014). Kramer et al. (2014) modelled the motion of the dust grain with considering two different cases. The first case, they use the Finson-Probstein model and neglect the comet gravity and the Lorentz force. The result shows that this model does not fit the observation data well (Kramer et al. 2014). The second case, they use the Lorentz force model with assuming the cometary nucleus as a sphere and considering the solar magnetic force and the solar gravity, the electric force, the solar radiation pressure, and the comet gravity are neglected, they found that this model is better for a dust grain with the size of $0.48 \sim 5.76 \mu m$ which is far away from the comet (Kramer et al. 2014). We now compare these different forces. The solar gravity force is calculated by

$$\mathbf{f}_{Sg} = \frac{GM_S m}{R^2} \frac{\mathbf{R}}{R}, \qquad (17)$$

where $M_S$ represent the mass of the Solar, $m$ represent the mass of the dust grain, $\mathbf{R}$ represent the position of the dust grain relative to the Solar, $R$ is the size of $\mathbf{R}$. The solar radiation pressure is calculated by

$$\mathbf{f}_{Sr} = \frac{K_r P_r A_r}{m} \frac{\mathbf{R}}{R}, \qquad (18)$$

where $K_r$ is the absorption coefficient, $P_r$ is the intensity of the solar radiation pressure, $A_r$ is the sectional area of the dust grain. Denote the area-mass ratio $a_m = \dfrac{A_r}{m}$.

The gas drag force is calculated by

$$\mathbf{f}_{gd} = \frac{1}{2} C_d \rho_{gd} A_r v_{gd}^2, \qquad (19)$$

where $C_d=1.0$ is the gas drag coefficient, $\rho_{gd}$ is density of the gas drag, $v_{gd}$ is the relative velocity between the dust grain and the gas frag. Other force models are presented in the previous sections, including the gravity force caused by the irregularly shaped cometary nucleus as well as the electric and magnetic force.

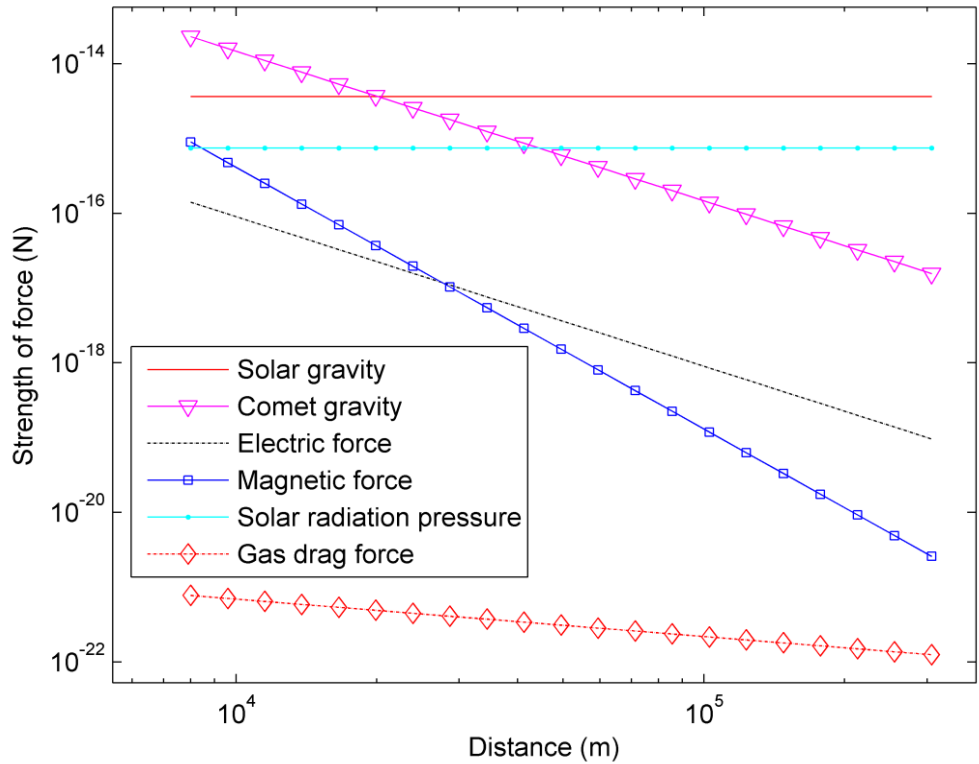

Fig. 7.a. Comparison of different forces on the dust grain: The mass ($1.0 \times 10^{-10}$kg), charge ($1.0 \times 10^{-16}$C), and sectional area ($2.0 \times 10^{-14}$m$^2$) are fixed, the distance of the dust grain relative to the mass centre of the cometary nucleus is changing.

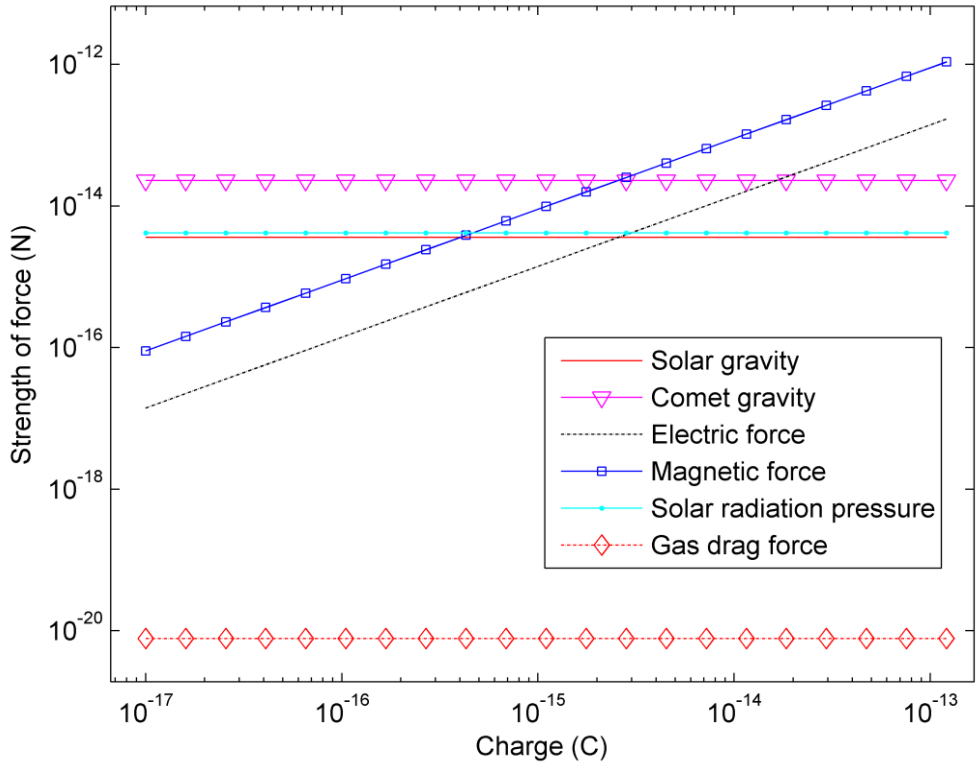

Fig. 7.b. Comparison of different forces on the dust grain: The mass ($1.0 \times 10^{-10}$ kg), sectional area ($2.0 \times 10^{-14}$ m$^2$), and position (relative to the mass center of the comet: 21km) are fixed, the charge is changing.

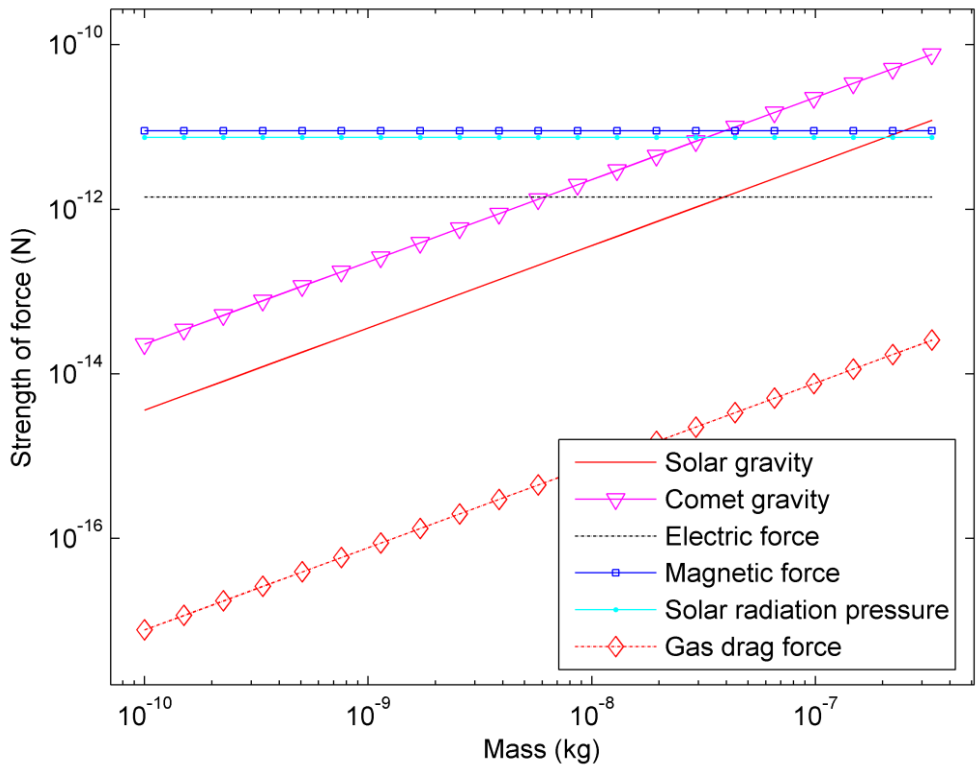

Fig. 7.c. Comparison of different forces on the dust grain: The charge ($1.0 \times 10^{-16}$ C), sectional

area($2.0 \times 10^{-14} \text{m}^2$), and position are (relative to the mass center of the comet: 21km) fixed, the mass is changing.

At first, we fix the mass, charge, and sectional area of the dust grain, let the distance of the dust grain relative to the mass centre of the cometary nucleus change. The initial values are $Q = 1.0 \times 10^{-16} C$, $K_r = 0.8$, $P_r = 4.65 \times 10^{-9} \text{N} \cdot \text{m}^{-2}$, $\rho_{gd} = 5.7 \times 10^{-5} \text{kg} \cdot \text{m}^{-3}$, and $a_m = 2.0 \times 10^{-7} \text{m}^2 \cdot \text{kg}^{-1}$. The charge of the comet is set to be $0.01C$. In this situation, we calculate the strength of different forces and show the result in Figure 7a. One can see that if the mass, charge, and sectional area are fixed, the solar gravity and solar radiation pressure acted on the dust grain are constants. When the dust grain is near the cometary nucleus, the comet gravity is the biggest one, and the gas drag force is the smallest one; however, the comet gravity and the electric and magnetic forces decrease rapidly while the distance is increasing.

Figure 7b shows the results of the strength of different forces when the charge is changing, and the mass, sectional area, as well as the position are fixed. The initial values are $r = 8.0 \text{km}$, $K_r = 0.6$, $P_r = 4.65 \times 10^{-9} \text{N} \cdot \text{m}^{-2}$, $\rho_{gd} = 5.7 \times 10^{-5} \text{kg} \cdot \text{m}^{-3}$, and $a_m = 2.0 \times 10^{-6} \text{m}^2 \cdot \text{kg}^{-1}$, where $r$ is the distance from the dust grain to the mass center of the cometary nucleus. In this situation, the electric and magnetic forces are changing. The comet gravity, the solar gravity, the solar radiation pressure, and the gas drag force are constants. When the value of the charge is small, the comet gravity is the biggest one; the solar gravity and the solar radiation pressure are both bigger than the Lorentz force, i.e. the electric and magnetic forces. When the value of the charge is big, the Lorentz force is the biggest one, and the solar gravity is the smallest one.

Figure 7c shows the results of the strength of different forces when the mass is changing, and the charge, sectional area, as well as the position are fixed. The initial values are $Q=1.0\times10^{-12}C$, $r=8.0\text{km}$, $K_r=0.8$, $P_r=4.65\times10^{-9}\text{N}\cdot\text{m}^{-2}$, $\rho_{gd}=5.7\times10^{-5}\text{kg}\cdot\text{m}^{-3}$, and $a_m=2.0\times10^{-6}\text{m}^2\cdot\text{kg}^{-1}$. In this situation, the solar radiation pressure and the Lorentz force are constants. The comet gravity and the solar gravity magnify while the mass magnify. When the value of the mass is small, the magnetic force and the solar radiation pressure are big, and the comet gravity and the solar gravity can be neglected. When the value of the mass is big, the comet gravity is the biggest one; other forces can be neglected in the sketchy calculation.

In Figure 7a-7c, the nucleus mass and the heliocentric distance have been used for all relevant parameters. The nucleus mass $m=2.4131\times10^{14}$ is calculated by the polyhedron method. From the above calculation, one can conclude that the motion of a dust grain depends on the area-mass ratio, the charge, and the distances of the dust grain to the mass center of the cometary nucleus and the Solar. The quantitative estimates of perturbing effects in this section show that the contents in Section 2 and Section 3 are suitable for large dust grains which are near the cometary nucleus and have small value of area-mass ratio. For a large dust grain (>1mm, more detailed can be seen in Ishiguro 2008) near the cometary nucleus with small value of area-mass ratio, the comet gravity is the major force acting on the dust grain. For a large dust grain far from the cometary nucleus with small value of area-mass ratio, the solar gravity and the solar radiation pressure are both major forces. For a small dust grain (<1mm, more detailed can be seen in Ishiguro 2008) near the cometary nucleus with large value of area-mass ratio, the solar radiation pressure and the comet gravity are two major forces. For a small dust grain near the cometary nucleus with large value of charge, the magnetic force, the solar radiation pressure, and the electric force are major forces. For a small dust grain far from the cometary nucleus, the solar gravity and solar radiation pressure are both major forces, and the comet gravity and comet Lorentz force can be neglected. The conclusions in Section 2 and Section 3 are suitable for a large dust grain near the cometary nucleus with small value of area-mass ratio and charge.

We now consider a dust grain which has a large area-mass ratio, consider the comet gravity force, Lorentz force, and the solar gravity as well as the solar radiation pressure. To compare the orbit with the orbit in the previous sections, we use the same orbital initial values in Fig. 6a. The orbital initial values is $\mathbf{r} = [953.6653\ -8897.3448\ 20.0065]\,\mathrm{m}$, $\mathbf{v} = [-0.0653020\ -0.00467089\ -0.01473039]\,\mathrm{m\cdot s^{-1}}$. The orbit is presented in Figure 8. The orbit showed in Fig. 6a is a stable periodic orbit, however, from Figure 8, one can see that if the dust grain has a large area-mass ratio, the dust grain will leave the comet and the orbit is no longer a periodic orbit. The orbit shape is helical while the dust grain leaves the comet, which coincides with the results in Kramer et al. (2014).

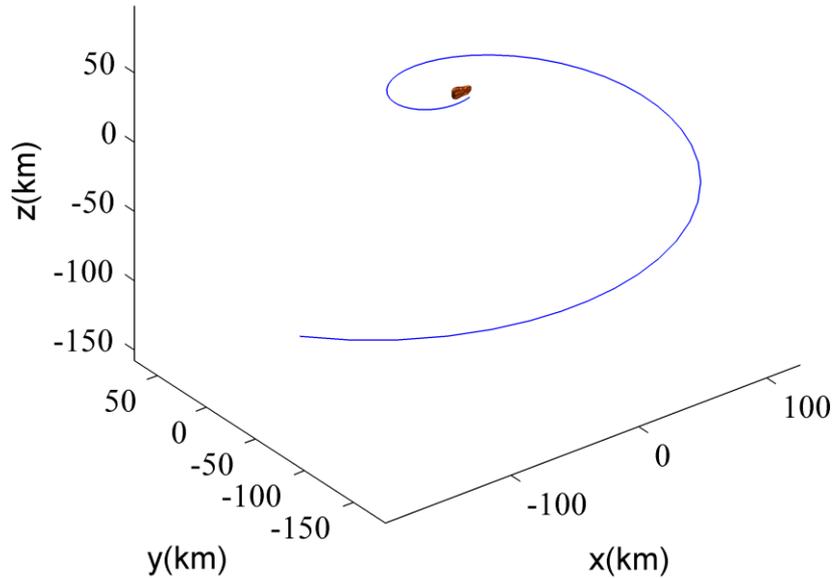

Fig. 8 The orbit of a dust grain which has a large area-mass ratio

## 5. Conclusions

Considering the gravitational field of an irregularly shaped body, as well as its electric and magnetic fields, the dynamical behavior of dust grains in the vicinity of a cometary nucleus is studied, including the local motion near equilibrium points, stable global motion, unstable global motion as well as resonant motion. We proved an identical equation for the number of non-degenerate equilibria of a dust grain in the combined gravitational and electromagnetic fields of a cometary nucleus. We found that the number of non-degenerate equilibria for the dust grain only varies in pairs, and the number of non-degenerate equilibrium points is an odd number. Besides, the degenerate equilibrium points of a dust grain may disappear, or may change to an arbitrary number (including zero) of degenerate equilibrium points and an even number (including zero) of non-degenerate equilibrium points; it is impossible that a degenerate equilibrium point changes to an odd number of non-degenerate equilibrium points.

The stability, bifurcations and resonances of periodic orbits for the dust grains in the irregular gravitational field, electric field, and magnetic field of a cometary nucleus are also discussed. The topological classification for stable periodic orbits and unstable periodic orbits is presented; conditions for bifurcations are also discussed; besides, resonant motion is also analyzed.

For the comet 1P/Halley's nucleus, there exist four equilibrium points outside the body, positions and eigenvalues of these equilibrium points can be computed. It is found that there are five topological classes of stable periodic orbits and six topological classes of unstable periodic orbits. There exist stable non-resonant periodic orbits, stable resonant periodic orbits and unstable resonant periodic orbits in the potential field of cometary nuclei.

The dust grain's motion depends on the area-mass ratio, the charge, and the distances of the dust grain to the mass center of the cometary nucleus and the Solar. The comet gravity is the major force acting on the dust grain if the dust grain is large and has small value of area-mass ratio. The solar gravity and the solar radiation pressure are the major forces if the dust grain is far away from the cometary nucleus. The comet gravity and the comet Lorentz force decrease rapidly while the distance from the dust grain to the mass center of the cometary nucleus is increasing.

**Acknowledgements**

This research was supported by the National Natural Science Foundation of China (No. 11372150), the Foundation of the State Key Laboratory of Astronautic Dynamics (No. 2017ADL-0202), and the National Basic Research Program of China (973 Program, 2012CB720000).

**Appendix A    Effective Potential and Equilibrium Points**

Defining the effective potential as

$$V(\mathbf{q}) = -\frac{1}{2}(\boldsymbol{\omega} \times \mathbf{q}) \cdot (\boldsymbol{\omega} \times \mathbf{q}) + U(\mathbf{q}) + Q\phi(\mathbf{q}) \tag{A1}$$

and substituting into Eq. (6) yields

$$\ddot{\mathbf{r}} + 2\boldsymbol{\omega} \times \dot{\mathbf{r}} + \frac{\partial V}{\partial \mathbf{r}} = Q\dot{\mathbf{r}} \times \mathbf{B}. \tag{A2}$$

The integral of the relative energy is in the form of

$$H = \frac{1}{2}\dot{\mathbf{q}} \cdot \dot{\mathbf{q}} + V(\mathbf{q}). \tag{A3}$$

Zero-velocity surfaces are given by the condition

$$V = H, \tag{A4}$$

so that the inequality $V(\mathbf{q}) > H$ defines forbidden regions for the motion of the dust grain, while $V(\mathbf{q}) < H$ defines allowed regions. Besides, the equality $V(\mathbf{q}) = H$ implies that the dust particle is static relative to the cometary nucleus.

The dynamical equations of the dust grain in the potential field of the nucleus can be expressed in Hamilton form as

$$\begin{cases} \dot{\mathbf{p}} = -\dfrac{\partial H}{\partial \mathbf{q}} \\ \dot{\mathbf{q}} = \dfrac{\partial H}{\partial \mathbf{p}} \end{cases}. \tag{A5}$$

The cometary nucleus rotates uniformly. We define the unit vector $\mathbf{e}_z$ by $\boldsymbol{\omega} = \omega \mathbf{e}_z$ and denote $\mathbf{v}_I = \dot{\mathbf{r}} + \boldsymbol{\omega} \times \mathbf{r}$, which is the velocity relative to the inertial frame. Then the mechanical energy of the dust grain $E = \frac{1}{2}\mathbf{v}_I \cdot \mathbf{v}_I + U(\mathbf{r})$ is not conserved, while the integral of the relative energy is conserved. The dynamical equations for the particle read in the body-fixed frame

$$\begin{cases} \ddot{x} - 2\omega\dot{y} + \dfrac{\partial V}{\partial x} + Q(\dot{z}B_y - \dot{y}B_z) = 0 \\ \ddot{y} + 2\omega\dot{x} + \dfrac{\partial V}{\partial y} + Q(\dot{x}B_z - \dot{z}B_x) = 0, \\ \ddot{z} + \dfrac{\partial V}{\partial z} + Q(\dot{y}B_x - \dot{x}B_y) = 0 \end{cases} \quad (A6)$$

where the effective potential is $V = U - \dfrac{\omega^2}{2}(x^2 + y^2) + Q\phi$. The integral of the relative energy is then expressed as

$$H = U + \frac{1}{2}(\dot{x}^2 + \dot{y}^2 + \dot{z}^2) - \frac{\omega^2}{2}(x^2 + y^2) + Q\phi = V + \frac{1}{2}(\dot{x}^2 + \dot{y}^2 + \dot{z}^2). \quad (A7)$$

The equilibrium points $(x_L, y_L, z_L)^T$ satisfy

$$\frac{\partial V}{\partial x} = \frac{\partial V}{\partial y} = \frac{\partial V}{\partial z} = 0. \quad (A8)$$

Then the linearised equations of motion relative to the equilibrium points can be expressed as

$$\begin{aligned} \ddot{\xi} - (2\omega + QB_z)\dot{\eta} + QB_y\dot{\zeta} + V_{xx}\xi + V_{xy}\eta + V_{xz}\zeta &= 0 \\ \ddot{\eta} + (2\omega + QB_z)\dot{\xi} - QB_x\dot{\zeta} + V_{xy}\xi + V_{yy}\eta + V_{yz}\zeta &= 0, \\ \ddot{\zeta} - QB_y\dot{\xi} + QB_x\dot{\eta} + V_{xz}\xi + V_{yz}\eta + V_{zz}\zeta &= 0 \end{aligned} \quad (A9)$$

where $\xi = x - x_L, \eta = y - y_L, \zeta = z - z_L$, and $V_{uv} = \left(\dfrac{\partial^2 V}{\partial u \partial v}\right)_L (u, v = x, y, z)$.

The characteristic equation of the equilibrium points is

$$\begin{vmatrix} \lambda^2 + V_{xx} & -(2\omega + QB_z)\lambda + V_{xy} & QB_y\lambda + V_{xz} \\ (2\omega + QB_z)\lambda + V_{xy} & \lambda^2 + V_{yy} & -QB_x\lambda + V_{yz} \\ -QB_y\lambda + V_{xz} & QB_x\lambda + V_{yz} & \lambda^2 + V_{zz} \end{vmatrix} = 0. \quad (A10)$$

Defining

$$\nabla^2 V = \begin{pmatrix} V_{xx} & V_{xy} & V_{xz} \\ V_{xy} & V_{yy} & V_{yz} \\ V_{xz} & V_{yz} & V_{zz} \end{pmatrix} = \begin{pmatrix} U_{xx} - \omega^2 + Q\phi_{xx} & U_{xy} + Q\phi_{xy} & U_{xz} + Q\phi_{xz} \\ U_{xy} + Q\phi_{xy} & U_{yy} - \omega^2 + Q\phi_{yy} & U_{yz} + Q\phi_{yz} \\ U_{xz} + Q\phi_{xz} & U_{yz} + Q\phi_{xy} & U_{zz} + Q\phi_{zz} \end{pmatrix} \quad (A11)$$

the following corollary gives a sufficient condition for the linear stability of the equilibrium points.

**Corollary 1**: If the matrix $\nabla^2 V$ is positive definite, the equilibrium point is linearly stable.

**Proof:**

Eq. (A9) can be rewritten as

$$\mathbf{M}\ddot{\boldsymbol{\rho}} + (\mathbf{G} + \widehat{\mathbf{B}})\dot{\boldsymbol{\rho}} + (\nabla^2 V)\boldsymbol{\rho} = 0, \tag{A12}$$

where $\boldsymbol{\rho} = [\xi \ \eta \ \zeta]^T$, $\mathbf{M} = \begin{pmatrix} 1 & 0 & 0 \\ 0 & 1 & 0 \\ 0 & 0 & 1 \end{pmatrix}$, $\mathbf{G} = \begin{pmatrix} 0 & -2\omega & 0 \\ 2\omega & 0 & 0 \\ 0 & 0 & 0 \end{pmatrix}$,

$\nabla^2 V = \begin{pmatrix} V_{xx} & V_{xy} & V_{xz} \\ V_{xy} & V_{yy} & V_{yz} \\ V_{xz} & V_{yz} & V_{zz} \end{pmatrix}$, and $\widehat{\mathbf{B}} = Q \begin{pmatrix} 0 & -B_z & B_y \\ B_z & 0 & -B_x \\ -B_y & B_x & 0 \end{pmatrix}$.

The matrices $\mathbf{M}$, $\mathbf{G}$, $\mathbf{K}_V$, and $\widehat{\mathbf{B}}$ satisfy $\mathbf{M}^T = \mathbf{M}$, $(\nabla^2 V)^T = \nabla^2 V$, $\mathbf{G}^T = -\mathbf{G}$, and $\widehat{\mathbf{B}}^T = -\widehat{\mathbf{B}}$, respectively.

Defining the Lyapunov function as

$$V_{Lyap} = \frac{1}{2}\left(\dot{\boldsymbol{\rho}}^T \mathbf{M} \dot{\boldsymbol{\rho}} + \boldsymbol{\rho}^T (\nabla^2 V) \boldsymbol{\rho}\right)$$

Then

$$\dot{V}_{Lyap} = \dot{\boldsymbol{\rho}}^T \left(\mathbf{M}\ddot{\boldsymbol{\rho}} + (\nabla^2 V)\boldsymbol{\rho}\right) = -\dot{\boldsymbol{\rho}}^T \mathbf{G} \dot{\boldsymbol{\rho}} - \dot{\boldsymbol{\rho}}^T \widehat{\mathbf{B}} \dot{\boldsymbol{\rho}} = 0$$

Because the matrices $\mathbf{M}$ and $\nabla^2 V$ are positive definite, we have $V_{Lyap} = \frac{1}{2}(\dot{\boldsymbol{\rho}}^T \mathbf{M} \dot{\boldsymbol{\rho}} + \boldsymbol{\rho}^T (\nabla^2 V)\boldsymbol{\rho}) > 0$, and therefore the equilibrium point is linearly stable. □

From the proof, it follows that if a dust grain's equilibrium point is linearly stable in the cometary nucleus' gravitational and electric fields, the existence and magnitude of the magnetic field has no effect to the stability of the equilibrium point.

## Appendix B  Proof of Theorem 1

**Proof:**

Let $\dot{\boldsymbol{\rho}} = \boldsymbol{\chi}$, substituting it into Eq. (A12) yields the following equation

$$\begin{bmatrix} \mathbf{I}_{3\times 3}\dot{\boldsymbol{\rho}} \\ \mathbf{M}\dot{\boldsymbol{\chi}} \end{bmatrix} = \begin{pmatrix} \mathbf{0}_{3\times 3} & \mathbf{I}_{3\times 3} \\ \nabla^2 V & -(\mathbf{G}+\hat{\mathbf{B}}) \end{pmatrix} \begin{bmatrix} \boldsymbol{\rho} \\ \boldsymbol{\chi} \end{bmatrix}. \tag{B1}$$

Defining $\boldsymbol{\Lambda} = \begin{bmatrix} \boldsymbol{\rho} \\ \boldsymbol{\chi} \end{bmatrix}$ Eq. (B1) is reduced to

$$\dot{\boldsymbol{\Lambda}} = \mathbf{g}(\boldsymbol{\Lambda}) = \mathbf{P}\boldsymbol{\Lambda}, \tag{B2}$$

where $\mathbf{g}(\boldsymbol{\Lambda})$ is a function of $\boldsymbol{\Lambda}$, and

$$\mathbf{P} = \begin{pmatrix} \mathbf{0}_{3\times 3} & \mathbf{I}_{3\times 3} \\ -\mathbf{M}^{-1}(\nabla^2 V) & -\mathbf{M}^{-1}(\mathbf{G}+\hat{\mathbf{B}}) \end{pmatrix}. \tag{B3}$$

From this expression we see that $\det \mathbf{P} = \det(\nabla^2 V)$.

If we define the function

$$f(\mathbf{r}) = \begin{pmatrix} \dfrac{\partial V}{\partial x} \\ \dfrac{\partial V}{\partial y} \\ \dfrac{\partial V}{\partial z} \end{pmatrix} + Q \begin{pmatrix} yB_z - zB_y \\ zB_x - xB_z \\ xB_y - yB_x \end{pmatrix} = \nabla V + Q\mathbf{r}\times\mathbf{B}, \tag{B4}$$

we have $\dfrac{df}{d\mathbf{r}} = \nabla^2 V$. Let $\Xi$ be the open set, use the topological degree theory (Mawhin and Willem, 1989, pages 116-119), then

$$\sum_{k=1}^{N}\left[\operatorname{sgn}\prod_{j=1}^{6}\sigma_j(E_k)\right] = \deg(f,\ \Xi,\ (0,\ 0,\ 0)) = const. \tag{B5}$$

So

$$\sum_{k=1}^{N}\left[\operatorname{sgn}\prod_{j=1}^{6}\sigma_j(E_k)\right] = \sum_{j=1}^{N}\left[\operatorname{sgn}(\det \mathbf{P})\right] = \sum_{j=1}^{N}\left[\operatorname{sgn}(\det(\nabla^2 V))\right] = const. \tag{B6}$$